\documentclass[sigconf]{acmart}
\def\plaintitle{Query Abandonment Prediction with Recurrent Neural Models of Mouse Cursor Movements}
\def\plainauthor{Lukas Brückner; Ioannis Arapakis; Luis A. Leiva}
\def\plainkeywords{Query Abandonment; Mouse Cursor Tracking; Deep Learning}

\usepackage[utf8]{inputenc}
\usepackage[FIGTOPCAP]{subfigure}
\graphicspath{{figures/}{plots/}}

\usepackage{array}
\newcolumntype{L}[1]{>{\raggedright\let\newline\\\arraybackslash\hspace{0pt}}m{#1}}
\newcolumntype{C}[1]{>{\centering\let\newline\\\arraybackslash\hspace{0pt}}m{#1}}
\newcolumntype{R}[1]{>{\raggedleft\let\newline\\\arraybackslash\hspace{0pt}}m{#1}}

\let\citet\cite

\copyrightyear{2020}
\acmYear{2020}
\setcopyright{acmlicensed}\acmConference[CIKM '20]{Proceedings of the 29th ACM International Conference on Information and Knowledge Management}{October 19--23, 2020}{Virtual Event, Ireland}
\acmBooktitle{Proceedings of the 29th ACM International Conference on Information and Knowledge Management (CIKM '20), October 19--23, 2020, Virtual Event, Ireland}
\acmPrice{15.00}
\acmDOI{10.1145/3340531.3412126}
\acmISBN{978-1-4503-6859-9/20/10}

\hypersetup{
  pdftitle={\plaintitle},
  pdfauthor={\plainauthor},
  pdfkeywords={\plainkeywords},
}

\begin{document}

\title{\plaintitle}

\author{Lukas Brückner}
\affiliation{
  \institution{Aalto University}
  \country{Finland}
}
\email{lukas.bruckner@aalto.fi}

\author{Ioannis Arapakis}
\affiliation{
  \institution{Telefonica Research}
  \country{Spain}}
\email{ioannis.arapakis@telefonica.com}

\author{Luis A. Leiva}
\affiliation{
  \institution{Aalto University}
  \country{Finland}
}
\email{firstname.lastname@aalto.fi}

\begin{abstract}
Most successful search queries do not result in a click if the user can satisfy their information needs directly on the SERP. Modeling query abandonment in the absence of click-through data is challenging because search engines must rely on other behavioral signals to understand the underlying search intent. We show that mouse cursor movements make a valuable, low-cost behavioral signal that can discriminate good and bad abandonment. We model mouse movements on SERPs using recurrent neural nets and explore several data representations that do not rely on expensive hand-crafted features and do not depend on a particular SERP structure. We also experiment with data resampling and augmentation techniques that we adopt for sequential data. Our results can help search providers to gauge user satisfaction for queries without clicks and ultimately contribute to a better understanding of search engine performance.
\end{abstract}

\ccsdesc[300]{Information systems~Search interfaces}
\ccsdesc[500]{Computing methodologies~Modeling and simulation}

\keywords{\plainkeywords}

\maketitle

\fancyhead{}

\section{Introduction}
It is no secret that users often abandon their searches.
However, most of these abandonments are deemed as \emph{good}
when the users' information needs are directly addressed
by the content that is available on the search engine results page (SERP).
Overall, the reasons for abandonment are well understood~\cite{Diriye12_abandonment}
and can help search engines to better estimate the success of a search session.
Nonetheless, a particular challenge arises in the absence of click-through data~\cite{Li09_abandonment, Huang11_clicks},
since search engines have to seek other behavioral signals
that can help explain the underlying user intent.

Traditionally, clicks and dwell times have been used as implicit feedback signals
to study user's search behavior on SERPs,
e.g., to predict search success~\cite{Hassan10_success}.
Conversely, the absence of clicks has been interpreted as a negative feedback signal of the quality of the results,
an assumption that has proved problematic~\cite{Joachims17}.
We argue that mouse cursor movements make a valuable, low-cost behavioral signal
that can tell good and bad abandonment apart.
To this end, we model mouse cursor movements on Yahoo SERPs using recurrent neural networks (RNNs)
that achieve competitive performance using sequences of $(x,y,t)$ coordinates as sole input.
We explore several architectural choices
and different data representations that do not rely on expensive hand-crafted features; c.f.~\cite{Arapakis16_kme}.
We also experiment with resampling and augmentation techniques that we adopt for sequential data.
Our findings can help search providers to better estimate user satisfaction for queries without clicks,
at scale, and ultimately contribute to a better understanding of search engine performance.
Our software is publicly available at \url{https://github.com/luksurious/abandonment-rnn}.

\section{Related Work}

Abandonment occurs when the searcher does not click on any of the links on the SERP,
including, e.g. advertisements, image carousels, cards, etc.
Some argue that clicking on a related search or spelling suggestion
should be regarded as an abandoned query,
since clicking on these links takes the user to another SERP~\cite{Diriye12_abandonment, Li09_abandonment}.
However, in this paper we focus exclusively on the cases where there is no click information at all,
thus making the abandonment prediction a much more challenging task than in previous work.

Historically, eye tracking has been used to understand user attention patterns on SERPs.
However, eye tracking requires specialized equipment
(ranging from expensive stationary eye trackers to more affordable but noisy webcams)
and so it is difficult to scale up to an online setting.
A large body of work has previously examined the relationship between eye gaze and mouse cursor movements during search tasks~\cite{Huang11_clicks, Guo11_success, Hassan10_success}, providing ample evidence on the connection between the two signals and highlighted mouse cursor tracking as a low-cost, scalable alternative to eye tracking.

Abandonment in web search has been widely used as a proxy of user satisfaction~\cite{Khabsa16_abandonment}.
Li et al.~\cite{Li09_abandonment} were the first to distinguish between good and bad abandonment
and motivated the need to augment click behavior to understand abandonment better.
Follow-up research took good abandonment into account~\cite{Arkhipova14_abandonment, Chuklin12_abandonment}.
Guo et al. engineered 30 features to predict document relevance~\cite{Guo12_dwell}
and 99 features to infer searcher's intent~\cite{Guo10_buy}.
Arapakis et al.~\cite{Arapakis16_kme} computed 638 features to predict user engagement on SERPs.
Finally, Diriye et al.~\cite{Diriye12_abandonment} used 2000 features to classify good and bad abandonment.

To the best of our knowledge, we are the first to predict query abandonment
with RNN models of mouse movements that do not use manually engineered features,
which are time-consuming and require domain expertise, see e.g.~\cite{Williams17_rnn}.
A key benefit of using RNNs is that they will pick up on features
that may not be evident yet important for prediction.

\section{Experiments}

Predicting good and bad abandonment from mouse movements
can be formulated as a binary sequence classification problem.
We show that this can be solved efficiently with RNNs that take \emph{raw} (unprocessed) mouse coordinates as input.
We also explore different data resampling and augmentation techniques for sequential data.

\subsection{Dataset}
\label{sec:dataset}

We use the dataset collected by Arapakis et al.~\cite{Arapakis16_kme},
which comprises 133 Yahoo SERPs browsed by 349 participants.
Each participant was given a question (e.g. ``How old is Brad Pitt?'') and an initial query (e.g. ``Brad Pitt age'')
that triggered a Knowledge Module (KM)~\cite{Arapakis15_kme}
on the SERP with information related to the query.
In many cases the answer to the question could be found in the KM,
but participants were allowed to reformulate the query or submit a new one.
Mouse cursor movements and additional metadata (e.g. user agent, screen size) were recorded.
After completing the search task,
users were asked if they noticed the KM (yes/no)
and the extent to which they found it useful (1--5 score).
Each participant was allowed to take part in the study only once.
Therefore, in this context, abandoned queries are those related to the last SERP
users interacted with and left with no clicks~\cite{Li09_abandonment, Huang11_clicks}.
A good abandonment is considered if users noticed the KM and found it useful (score $\geq 4$),
otherwise it is considered as a bad abandonment.

\begin{figure}[ht]
    \centering
    \def\w{0.9\linewidth}
    \subfigure[Distribution of dwell times\label{fig:dwell_times}]{\includegraphics[width=\w]{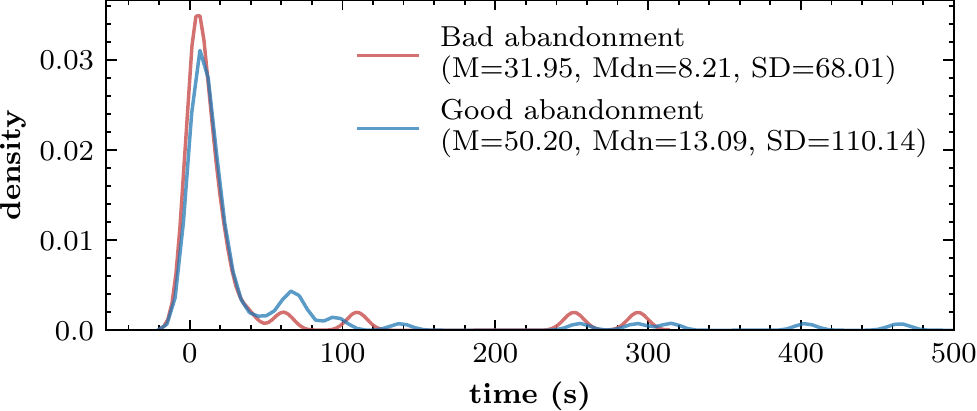}}\\[1em]
    \subfigure[Distribution of time offsets\label{fig:time_offsets}]{\includegraphics[width=\w]{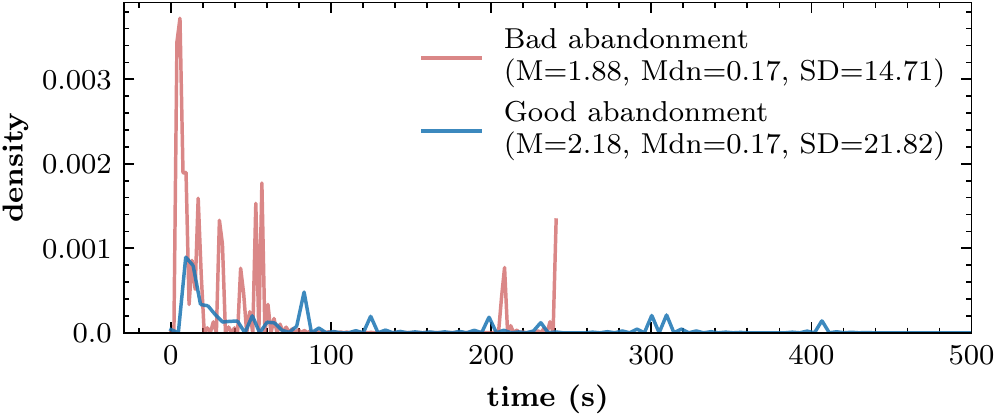}}
    \caption{Distribution of dwell times (top) and time offsets between consecutive mouse cursor movements (bottom).
      Approximately the same amount of time is spent on SERPs that led to either good or bad abandonment,
      though bad abandonments often comprise much shorter movements.}
    \label{fig:time-distr}
\end{figure}

The polling resolution of the \texttt{mousemove} events is 150\,ms~\cite{Arapakis16_kme}.
Mouse sequences with only one event, or timestep, were not considered for analysis.
And even though sequences with less than 5 timesteps can be considered very short,
we have observed that dwell times can be in the order of seconds (\autoref{fig:dwell_times}),
so they may still be informative.
We concluded to a dataset of 107 mouse sequences belonging to abandoned queries.
The majority of sequences has less than 30 \texttt{mousemove} events (M=25, Mdn=19, SD=22),
and only 10\% of the sequences have more than 50 timesteps.

\subsection{Model}

We use an RNN with bidirectional long short-term memory (BiLSTM),
which can handle long-term dependencies
and exploit both past and future information at every timestep.
To find the best architecture,
a non-exhaustive search was performed with different configuration options,
ranging from 1--3 recurrent layers with 25--100 units per layer.
We also tested the self-attention mechanism~\cite{Luong15}.
Hyperparameter search (e.g. learning rate, batch size, number of layers, units, etc.)
was performed with the Optuna framework,\footnote{\url{https://optuna.org/}}
monitoring the F-measure on the validation data.

The final architecture consists of two stacked BiLSTM layers with 100 units each and a dropout rate of 30\%.
Classification performance was reduced by $\sim$2\% in F-measure when implementing only one BiLSTM layer.
A similar performance degradation was observed with self-attention.
Adding a third recurrent layer did not significantly improve performance (less than 1\% of Precision and Recall).
Thus, given that model training stabilized in any case,
we favored the non-attentive architecture with two stacked BiLSTM layers.
Our RNN model takes as input a sequence of mouse cursor positions
and outputs the probability of good abandonment.

The input layer of our model is limited to 50 timesteps,
informed by our previous observations (\autoref{sec:dataset}).
Shorter sequences are padded with dummy values.
Longer sequences are truncated to the \emph{last} 50 timesteps,
as we argue that the decision to leave the SERP happens rather towards the end of the browsing session
and thus any meaningful interaction that may signal a good or bad abandonment
is most likely to be found in the later movements.
Previous research has found that user's decision process
is initiated unconsciously at first and enter consciousness afterward~\cite{Haggard11}.

\subsection{Data Representation}

We explore several lightweight data representation formats,
starting with \emph{raw} sequences of $(x,y)$ mouse coordinates.
This representation, however, ignores movement speed and time elapsed between consecutive coordinates,
which may also contain interesting behavioral information.
Thus, we also experiment with temporal information:
We compute \emph{time offset} as the difference between consecutive timestamps (in ms)
and \emph{speed} as the Euclidean distance between consecutive coordinates divided by the time offset.
As shown in \autoref{fig:time_offsets}, time offsets can range from a few seconds to six minutes.
We can see that mouse sequences belonging to the bad abandonment class
have a higher density at very short time ranges, suggesting a wandering browsing behavior,
i.e. with no intent or a clear goal in mind~\cite{MartinAlbo16_insight}.

Another variation we consider is that of standardizing the mouse coordinates to a common screen size,
since the dataset was collected remotely and therefore screen sizes vary across participants.
Given the left-aligned layout design of the SERP,
horizontal coordinates are scaled to a common screen width of 1280\,px.

\subsection{Data Resampling and Augmentation}
\label{sec:augmentation}

\begin{figure}[!ht]
    \centering
    \includegraphics[width=\linewidth]{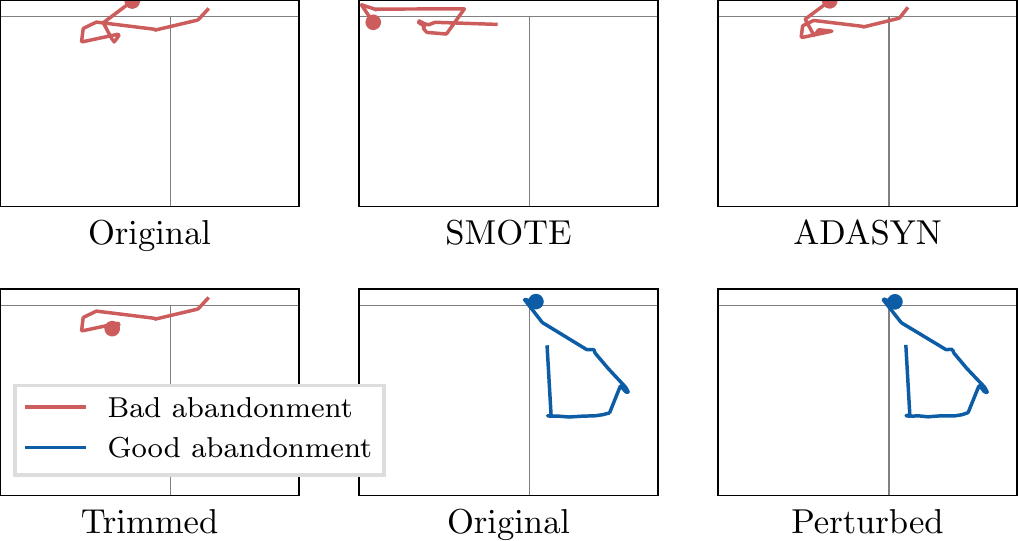}
    \caption{Examples of original and augmented mouse sequences.
        Gray lines depict the SERP skeleton: search bar (top-most rectangle)
        and the KM component (right-most rectangle).
        The dots indicate the initial mouse coordinate.}
    \label{fig:augmented-seqs}
\end{figure}

In our dataset, 30 instances belong to the negative class (bad abandonment)
and 77 to the positive class (good abandonment).
Other researchers have reported similar ratios in previous work~\cite{Stamou10_inactivity}.
Because of this small-sized imbalanced dataset, training leads to a degenerate model
that essentially predicts the majority class,
even if we compensate training with class weighting.
To address this issue, we test different data resampling techniques.
Undersampling the majority class does not perform well because valuable training data is ignored.
On the contrary, by oversampling the minority class we end up with an augmented and balanced training set.
We use the SMOTE~\cite{Chawla_2002} and ADASYN~\cite{ADASYN} oversampling techniques
on the training partition only,
after creating all splits for testing and validation, to prevent data leakage.
These techniques generate new point coordinates by interpolating among all available training sequences (see \autoref{fig:augmented-seqs} for some examples).

Further, we experiment with augmenting the training data
using domain knowledge,
for which we apply a random uniform distortion of 0--2\,px
to each mouse coordinate.
This distortion is aimed to imitate the original mouse cursor movements
but with slightly different positions,
inspired by previous work that examined user's pointing behavior~\cite{MartinAlbo16_insight}.
We also experiment with randomly trimming the mouse sequences
by up to 5 timesteps at the beginning of each sequence.
In total, we tested 4 different data augmentation techniques:
(1)~only distortion, (2)~only trimming, (3)~distortion followed by trimming,
and (4)~either distortion or trimming (but not combined).
After data augmentation, the training partition contained 115--140 sequences per class.
Since both classes were augmented, a perfectly balanced dataset is achieved
by removing augmented sequences from the majority class at random.

\subsection{Model Training}
\label{sec:training}

\begin{table*}[!ht]
\centering

  \begin{tabular}{*3l c *4r}
  \toprule

  \textbf{Input data}
    & \textbf{Time}
    & \textbf{Augmentation}
    & \textbf{Adj. Precision}
    & \textbf{Adj. Recall}
    & \textbf{F-measure}
    & \textbf{ROC AUC} \\

  \midrule

  \multicolumn{3}{l}{\textit{All abandoned queries are considered bad abandonments}}
    & 0.08 [0.06, 0.11]
    & 0.28 [0.24, 0.32]
    & 0.12 [0.10, 0.15]
    & 0.50 [0.50, 0.50] \\

  \multicolumn{2}{l}{RF using 10-dim feat vectors}
    & ADASYN
    & 0.67 [0.58, 0.76]
    & 0.64 [0.54, 0.73]
    & 0.64 [0.54, 0.73]
    & 0.60 [0.50, 0.69] \\

  \multicolumn{2}{l}{XGB using 10-dim feat vectors}
    & ADASYN
    & 0.70 [0.60, 0.78]
    & 0.65 [0.55, 0.74]
    & \textbf{0.65} [0.55, 0.74]
    & 0.61 [0.51, 0.70] \\

  \midrule

  Standardized coords
    & no
    & none
    & 0.52 [0.48, 0.57]
    & \textbf{0.72} [0.68, 0.76]
    & 0.60 [0.56, 0.65]
    & 0.50 [0.46, 0.55] \\

  Raw coords
    & yes
    & rand. undersample
    & 0.68 [0.64, 0.72]
    & 0.59 [0.54, 0.63]
    & 0.59 [0.54, 0.63]
    & 0.59 [0.54, 0.63] \\

  Standardized coords
    & yes
    & rand. oversample
    & 0.67 [0.62, 0.71]
    & 0.63 [0.58, 0.67]
    & 0.62 [0.57, 0.66]
    & 0.59 [0.54, 0.63] \\

  Speed only
    & implied
    & SMOTE
    & 0.67 [0.63, 0.71]
    & 0.63 [0.58, 0.67]
    & 0.63 [0.58, 0.67]
    & 0.59 [0.55, 0.63] \\

  Speed + distance to KM
    & implied
    & SMOTE
    & 0.70 [0.65, 0.73]
    & 0.65 [0.61, 0.69]
    & \textbf{0.65} [0.61, 0.69]
    & 0.61 [0.57, 0.65] \\

  Raw coords
    & yes
    & SMOTE
    & 0.69 [0.65, 0.73]
    & 0.63 [0.59, 0.67]
    & 0.63 [0.59, 0.68]
    & 0.61 [0.57, 0.65] \\

  Standardized coords
    & yes
    & ADASYN
    & 0.68 [0.64, 0.72]
    & 0.64 [0.60, 0.68]
    & 0.64 [0.59, 0.68]
    & 0.61 [0.56, 0.65] \\

  Standardized coords
    & yes
    & distortion or trimming
    & \textbf{0.72} [0.68, 0.76]
    & 0.65 [0.61, 0.69]
    & \textbf{0.65} [0.61, 0.69]
    & \textbf{0.63} [0.59, 0.68] \\

  \bottomrule

  \end{tabular}

  \caption{Experiment results.
    Top rows are baseline conditions.
    We report the best combination of \{Coords, Time, Resampling, Augmentation\}
    techniques tested (36 in total).
    95\% conf. intervals according to the Wilson method for binomial distributions.
  }
  \label{tab:results}
\end{table*}

We use nested 10$\times$5-fold stratified cross-validation (CV):
The dataset is split into 10 equal parts in the outer CV loop,
with each part being once used as a test set (10\%),
maintaining the class balance of the original data in each split.
The remaining 90\% of the data is split into 5 equally sized parts in the inner CV loop,
with one part being used as a validation set and four parts as training data.
Nested CV helps to better estimate the generalization capabilities of our models
and avoids potential biases from evaluating on a single split of the data.
The model is trained in batches of size 4
with the popular Adam optimizer ($\eta=0.0001, \beta_1=0.9, \beta_2=0.999$)
up to 100 epochs using binary cross-entropy loss
We use early stopping if the F-measure on the validation set does not improve in 5 consecutive epochs,
and retain the learned weights according to the best F-measure.

\subsection{Baseline Models}

We compare our models against several baselines,
starting with the usual assumption
in which \emph{all} abandoned queries are considered bad abandonments.
We should remind the reader that the goal of this paper is to investigate models
that (1)~do not require expensive manual feature engineering
and (2)~do not depend on a particular SERP structure.
Therefore, we implement random forest (RF) and extreme gradient boosting (XGB) classifiers
using simple features derived from client-side interactions,
informed by previous work~\cite{Arapakis16_kme, Diriye12_abandonment, Guo10_buy, Guo12_dwell, Leiva13_smt}:
dwell time, avg. time offset, num. mousemoves, num. hovers over the KM, num. scrolls, trajectory length, range, and scroll reach.
Both range and scroll reach refer to both vertical and horizontal components,
so we use 10-dim feature vectors for our RF and XGB classifiers.
Finally, we also train an RNN model using mouse speed and distance to the KM at each timestep.

\subsection{Results}
\label{sec:results}

In \autoref{tab:results} we report the weighted macro-averages of Precision, Recall, F-measure, and Area Under the ROC Curve (AUC)
over the test partition, which simulates unseen data.
For brevity's sake, we only report the results of the best experiment configuration found
for each data representation, resampling, and augmentation techniques.
Note that we tested 36 configuration combinations in total,\footnote{
  We considered
  \{raw, standardized\} coords
  $\times$ \{time, no time\} information
  $\times$ \{weighted, undersampling, oversampling, SMOTE, ADASYN,
  distortion only, trimming only, distortion followed by trimming, either distortion or trimming\} augmentation.
}
and in many cases the differences between combinations are small.
The top rows in the table denote the baseline models.

We can see that the usual assumption of considering abandoned queries as bad abandonments
is not useful to search engines. In addition, it has no discriminative power, as indicated by the AUC score.
The RNN trained on non-augmented data achieved the best Recall, as expected,
since it is biased to predicting the majority class, sacrificing Precision and AUC as well.
Our results show that adding temporal information leads to a better classification performance.
The tested resampling techniques (SMOTE and ADASYN) improved performance further,
as they bring in more data for analysis.
Resampling was also beneficial for the baseline models (RF and XGB).
We found that standardizing the mouse coordinates improves RNN performance further and outperforms the baseline models.
As shown in the last row of \autoref{tab:results},
the best overall result was achieved with the ``distortion or trimmed''
data augmentation technique
together with standardized coordinates and time offsets as input to the RNN.

\section{Discussion and Future Work}

Our experiments illustrate that it is possible to use RNNs to tell good and bad abandonment apart
without having to engineer sophisticated features,
lowering thus the barrier to creating more expressive user interaction models.
While we trained several models using F-measure as monitoring metric, thereby balancing Precision and Recall,
in a commercial setting Precision might be more important than Recall,
or vice versa, and thus it would make sense to optimize the model for either metric.
For example, if a search engine wants to focus on identifying a good abandonment correctly,
then it should prime Precision over Recall.

Our model can predict good and bad abandonment in the absence of click-through data,
though the underlying reasons of abandonment can vary.
These include, e.g. satisfaction or dissatisfaction with the results, interruptions, or undirected searching.
Researchers have discussed the many challenges associated to this label acquisition task~\cite{Liu15_opinions, Huang11_clicks}.
In addition, cognitive decision-making is a short-term process
so there may be a limited range of mouse cursor patterns associated with those processes.
Ultimately, being able to accurately predict good and bad abandonment has implications for search engine design and evaluation.
For example, our model predictions can be used to complement click-based metrics of search performance
by e.g. assigning an abandonment rate to the query or estimating the type of abandonment (i.e. good or bad) that occurred.

The sample size of our user study is relatively small
and therefore a large-scale study replication is needed as future work.
Nevertheless, through our experiments, we have demonstrated that we can
train compelling RNN models that do not require expensive feature engineering
and, most importantly, do not depend on SERP contents and/or a particular layout structure.
Recent work has shown that neural models of mouse movements
can deliver competitive results with little training data~\cite{Arapakis20_mdtl}.
Taken together, our contributions improve our understanding of search engine performance
and can help search providers to better estimate user satisfaction for queries without clicks.

\begin{small}
\vspace{1em}
\noindent\textbf{Acknowledgments:}
We thank Jeff Huang for reviewing an earlier version of this paper.
\end{small}

\end{document}